\documentclass[prd,twocolumn,preprintnumbers,floatfix]{revtex4-1}

\usepackage{amsmath}
\usepackage{epsfig}
\usepackage{graphicx}
\usepackage{color}
\usepackage[normalem]{ulem}
 \usepackage{url}
\usepackage[breaklinks, plainpages=false, colorlinks=true, anchorcolor=cyan, linkcolor=red, citecolor=cyan, urlcolor=magenta, bookmarks=false]{hyperref}

\usepackage[caption=false]{subfig}

\begin{document}
\renewcommand{\thefigure}{\arabic{figure}}
\setcounter{figure}{0}

 \def\I{{\rm i}}
 \def\E{{\rm e}}
 \def\D{{\rm d}}

\bibliographystyle{apsrev}

\title{Modeling non-stationary noise: applications in gravitational wave astronomy}

\author{Neil J. Cornish}
\affiliation{eXtreme Gravity Institute, Department of Physics, Montana State University, Bozeman, Montana 59717, USA}

\begin{abstract} 
In an ideal world, the measurement noise in gravitational wave data would be stationary and Gaussian. In reality, neither of these conditions holds. Here a general framework is introduced that can be used to model non-stationary noise in an easily interpretable way, using a dynamic power spectrum $S(f,t)$. The construction is a Gram-factor model for the noise covariance matrix that is positive semi-definite by construction. This construction generalizes the familiar stationary power spectrum $S(f)$. The dynamic spectrum encodes the properties of the noise covariance matrix in any basis, including the frequency domain, time domain, and time-frequency wavelet domain. Closed form expressions are given for discrete Fourier representations of the data, and for discrete Wilson-Daubechies wavelet representations of the data. Both take the form of Gramian matrices. Examples are provided, including the non-stationarity caused by window functions, the modulated response to galactic binary signals for space-based detectors, and other, more general types of non-stationarity.
\end{abstract}

\maketitle

\section{Introduction}

To extract information about gravitational wave signals we need accurate models of the noise, as the noise model defines the likelihood function that is at the heart of all gravitational wave analyses~\cite{Finn:1992wt}. The data produced by gravitational wave detectors is a time series, and standard time-series analysis techniques are used to separate signals from noise. In most analyses, the assumption is made that the data are stationary and Gaussian, but in reality this is not always the case~\cite{LIGOScientific:2019hgc}. Techniques have been introduced to mitigate the effects of instrumental transients, or glitches~\cite{Cornish:2014kda,Cornish:2020dwh,Chatziioannou:2021ezd,Hourihane:2022doe}, which can make the data non-Gaussian, but less work has been done to account for time variation in the overall noise properties. It has been shown that non-stationary noise can impact searches for binary mergers~\cite{Zackay:2019kkv,Mozzon_2020}, the detectability of massive black hole binaries and galactic binaries~\cite{Digman:2022jmp}, and estimates of the Hubble parameter~\cite{Mozzon:2021wam}.

Here a simple and easily interpretable generalization of the familiar power spectrum, $S(f)$, which is used in both ground-based interferometer analyses and pulsar timing array analyses, is proposed. An operational definition of the dynamic spectrum $S(f,t)$ is given that generalizes the notion of a stationary power spectrum to cover noise processes that have properties that vary in time. This dynamic spectrum completely defines the noise covariance matrix for discretely sampled, finite duration data. This simple definition ignores the additional inter-frequency correlations that come from truncating a continuous, infinite-in-time noise process. But it can be shown that these additional correlations are negligible unless the noise has an extremely steep dependence on frequency. Note that these additional correlations are important for the lowest frequencies probed by pulsar timing arrays~\cite{Allen:2024uqs,Crisostomi:2025vue}, but can be safely ignored for ground-based and future space-based detectors as they only impact frequencies that are outside the useful sensitive band of the detectors. In a follow-up paper~\cite{ghoshcornish2026} we introduce a method to estimate $S(f,t)$ from real gravitational wave data, and explore the impact that non-stationarity has on parameter estimation for astrophysical sources.

This construction is a Gram-factor model for the covariance matrix. Positivity of the spectrum is enforced by setting $S(f,t) = s^2(f,t)$. The amplitude functions $s(f,t)$ define the covariance square root, so the resulting covariance is positive semidefinite by construction. The covariance matrix is then Gramian by construction, as we will see from the explicit expressions for the Fourier domain and wavelet domain covariance matrices. The Gramian structure allows for efficient calculation of the covariance, and for a straightforward method to simulate non-stationary data.

\section{Non-stationary noise}

There have been many attempts to define dynamic spectra, the most well known being the Wigner-Ville spectrum~\cite{Wigner:1932,Ville:1948,MartinFlandrin:1985} $S_{\rm WV}(f,t)$. For a continuous in time, infinite in duration, real noise process $x(t)$, the Wigner-Ville spectrum can be constructed as the Fourier transform of the symmetrized noise correlation function:
\begin{equation}
S_{\rm WV}(f,t) = \int_{-\infty}^{\infty} d \tau e^{-2\pi i f \tau} {\rm E}[ x(t + \tau/2) x(t-\tau/2) ] \, .
\end{equation}
The Wigner-Ville spectrum is related to the non-stationary noise correlation function via the Wiener-Khinchin theorem:
\begin{equation}
C(\tau, t) =  \int_{-\infty}^{\infty} d f e^{2\pi i f \tau}  S_{\rm WV}(f,t) \, .
\end{equation}
In the limit that the spectrum is stationary, these expressions reduce to their familiar forms. Unfortunately, despite this nice formulation, the Wigner-Ville spectrum does not define a true power spectrum as it can take negative values. More accurately, it is a local covariance representation that has the right marginals and the right stationary limit, but the bilinear construction produces interference terms, so it can go negative. Various approaches have been proposed to correct this deficiency~\cite{lu2009wienerkhinchintheoremnonwidesense,Dechant_2015}, the simplest being to introduce an additional average over some time interval centered on time $t$. Here we take a more direct approach and define $S(f,t) = s(f,t)^2$ as a positive dynamic spectrum. For finite observation times $T$, the main result of this paper is that the noise covariance matrix in the discrete Fourier domain, $C_{ml} = {\rm E}[ {\tilde x}[m] {\tilde x}^*[l]]$,  is given by
\begin{equation}\label{main}
C_{ml} = \frac{1}{N}  \sum_{k=0}^{N-1}  \tilde{s}_k[\{m-k\}_N]   \tilde{s}_k^*[\{l-k\}_N]\, .
\end{equation}
where ${\tilde x}[m]$ is the discrete Fourier transform of $x(t)$, and $s_k[m]$ is the discrete Fourier transform of $s(f_k,t)$, where $f_k = k/T$. We have used the notation $\{j\}_N = (j) {\rm mod}\, N$. The Fourier noise covariance matrix is Gramian, and hence Hermitian, $C_{ml} = C^*_{lm}$. The imaginary parts of the noise covariance matrix encode information about time. The full derivation is given in section~\ref{model}. 

With the Fourier domain noise covariance matrix in hand, it is a straightforward calculation to find the noise covariance matrix in any other representation, since they are related by linear transformations. If $y=L\tilde{x}$, then $C_y=LCL^\dag$. For example, the time domain noise covariance matrix $C_T$ and the wavelet domain covariance matrix $C_W$ are given by
\begin{equation}
C_{T} = F^{-1} C (F^{-1})^\dag, \quad
C_{W} = W F^{-1} C (F^{-1})^\dag W^T \, .
\end{equation}
Here $F$ is the unnormalized discrete Fourier transform matrix and $W$ is the real discrete wavelet transform matrix. The inverse Fourier matrix obeys $F^{-1}=F^\dag/N$, while the wavelet matrix is orthogonal, $W^{-1}=W^T$. The operation $\tilde{x} = F x$ on a column vector $x$ corresponds to performing a discrete Fourier transform. This order $N^2$ operation can be reduced to order $N\log N$ using a fast Fourier transform. Similarly, if $W$ is a Wilson-Daubechies (WD) wavelet transformation~\cite{Daubechies:1991wv,Necula_2012}, the transform can also be performed using fast Fourier transforms. That reduces the computational cost of the matrix operations from $N^3$ to $N^2 \log N$.

It is worth emphasizing that the noise weighted inner product, which plays a prominent role in gravitational wave data analysis, is invariant under changes in basis. Writing the inner product as
\begin{equation}
(a|b) = a^\dag C^{-1} b
\end{equation}
and changing variables with any invertible matrix $L$, $a'=La$, $b'=Lb$, and $C'=LCL^\dag$, we find
\begin{eqnarray}
(a|b) &=& a^\dag C^{-1} b \nonumber \\
&= & (L a)^\dag ( L C L^\dag)^{-1} (L b)  \nonumber \\
& =  & {a'}^\dag  {C'}^{-1} b' \, .
\end{eqnarray}
Information is only lost if a diagonal approximation for the noise covariance is adopted when it is not warranted, such as with non-stationary noise.

The other main result of this paper is an explicit expression for the WD-wavelet noise covariance matrix,  $S_{ij} = {\rm E}[w_i w_j]$, where $i=(m_i,n_i)$ labels a time-frequency pixel. Here $w_i$ is the real amplitude of the wavelet pixel at time index $n_i$ and frequency index $m_i$. Using the Fourier domain noise covariance and the Fourier to WD-transformation introduced in Ref.~\cite{Cornish:2020odn}, we first define
\begin{equation}
K_{ij} = \frac{1}{N}
\sum_{k=0}^{N-1} U_{ki}U^*_{kj}, \quad
H_{ij} = \frac{1}{N}
\sum_{k=0}^{N-1} U_{ki}V^*_{kj} \, ,
\end{equation}
where
\begin{eqnarray}
U_{ki} &=& \sum_l Q_i(l)\Phi[l]\,
\tilde{s}_k[\{a_i(l)-k\}_N] \, , \nonumber\\
V_{ki} &=& \sum_l Q_i^*(l)\Phi[l]\,
\tilde{s}_k[\{-a_i(l)-k\}_N] \, ,
\end{eqnarray}
and
\begin{equation}
a_i(l)=l+\frac{m_iN_t}{2}, \quad
Q_i(l)=\exp\left(\frac{2\pi i(l+N_t/2)n_i}{N_t}\right) .
\end{equation}
The exact real WD covariance is
\begin{equation}
S_{ij} =
{\cal A}_f^2\left(
\beta_i\beta_j^*K_{ij}
+\beta_i^*\beta_j K_{ji}
+\beta_i\beta_jH_{ij}
+\beta_i^*\beta_j^*H_{ij}^*
\right) ,
\end{equation}
where the coefficients $\beta_i$ implement the even and odd WD parity projections: $\beta_i=1/2$ for even $n_i+m_i$ and $\beta_i=(-1)^{m_i}/(2\I)$ for odd $n_i+m_i$. We see that the WD covariance is built from Gramian matrices of the amplitude functions $s(f,t)$ after they have been filtered by the WD windows $\Phi[l]$. Note the similarity between the wavelet domain and Fourier domain noise covariance matrices. Both involve sums over the Fourier transform of the amplitude functions, with the wavelet version picking up a window function. For a comprehensive review of the popular Wilson-Daubechies-Meyer (WDM) transform see Refs.~\cite{Johnson:2026rrn,Vajpeyi:2026msr}.

\section{Modeling non-stationary noise\label{model}}

\subsection{Preliminaries: Modulated stationary noise processes}

Consider a finite duration of time, $T$, and evenly sampled data with cadence $dt=T/N$ and $N$ samples. Suppose that $x(t)$ is a stationary noise process that gets multiplied by a modulation $g(t)$ to give the non-stationary process $y(t) = g(t) x(t)$. We adopt the discrete Fourier transform (DFT) convention 
\begin{equation}
{\tilde x}[k]=  \sum_{n=0}^{N-1} x[n] e^{-2\pi i kn/N}
\end{equation}
with the inverse DFT carrying a factor of $1/N$. For stationary noise with expectation values:
\begin{eqnarray}
{\rm E}[{\tilde n}[j]] &=& 0 \nonumber \\
{\rm E}[{\tilde n}[j] {\tilde n}^*[k]] &=& N S[k] \delta_{jk} 
\end{eqnarray}
we find that the Fourier domain noise covariance matrix $C_{ml}={\rm E}[ {\tilde y}[m] {\tilde y}^*[l]]$ is 
\begin{equation}\label{cov}
C_{ml}= \frac{1}{N} \sum_{k=0}^{N-1} S[k] \tilde{g}[\{m-k\}_N]   \tilde{g}^*[\{l-k\}_N] \, .
\end{equation}
The matrix $C$ is Hermitian, $C=C^\dag$, with the complex terms encoding the time dependence caused by the modulation.
Note that for $g(t)=1$ we have ${\tilde g}[j] = N \delta_{0j}$ and $C_{ml} = N S[m] \delta_{ml}$, which is the usual result for stationary noise. One application of this result that is relevant to gravitational wave astronomy is when $g(t)$ is an apodizing filter, such as a Hann or Tukey window, which is used to suppress spectral leakage when analyzing finite stretches of a longer time series~\cite{Talbot:2021igi,Talbot:2025vth}. Because $g(t)$ is not flat, the windowed data $y(t)$ become non-stationary. 

\begin{figure}[htp]
\includegraphics[width=0.45\textwidth]{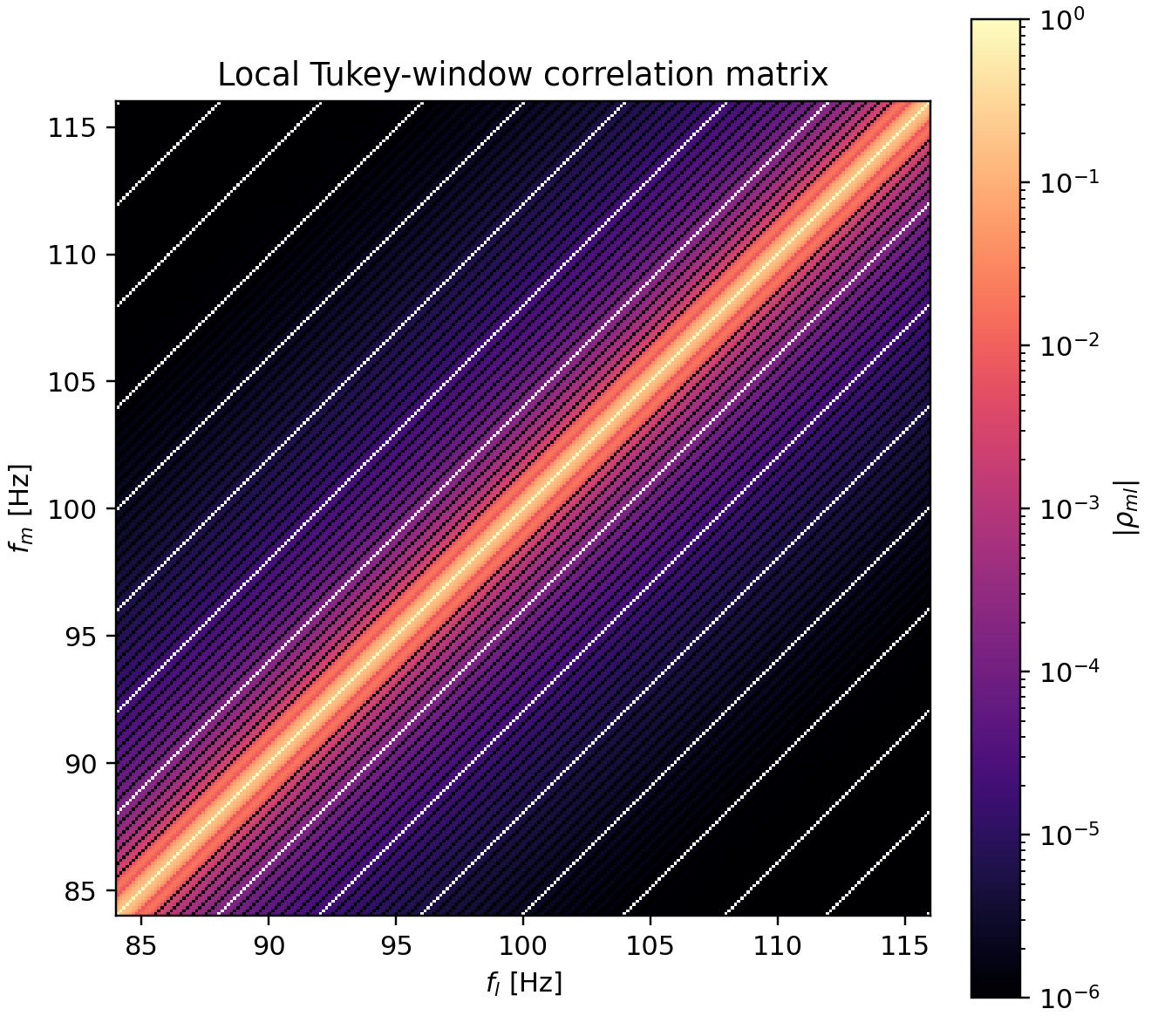} 
\includegraphics[width=0.45\textwidth]{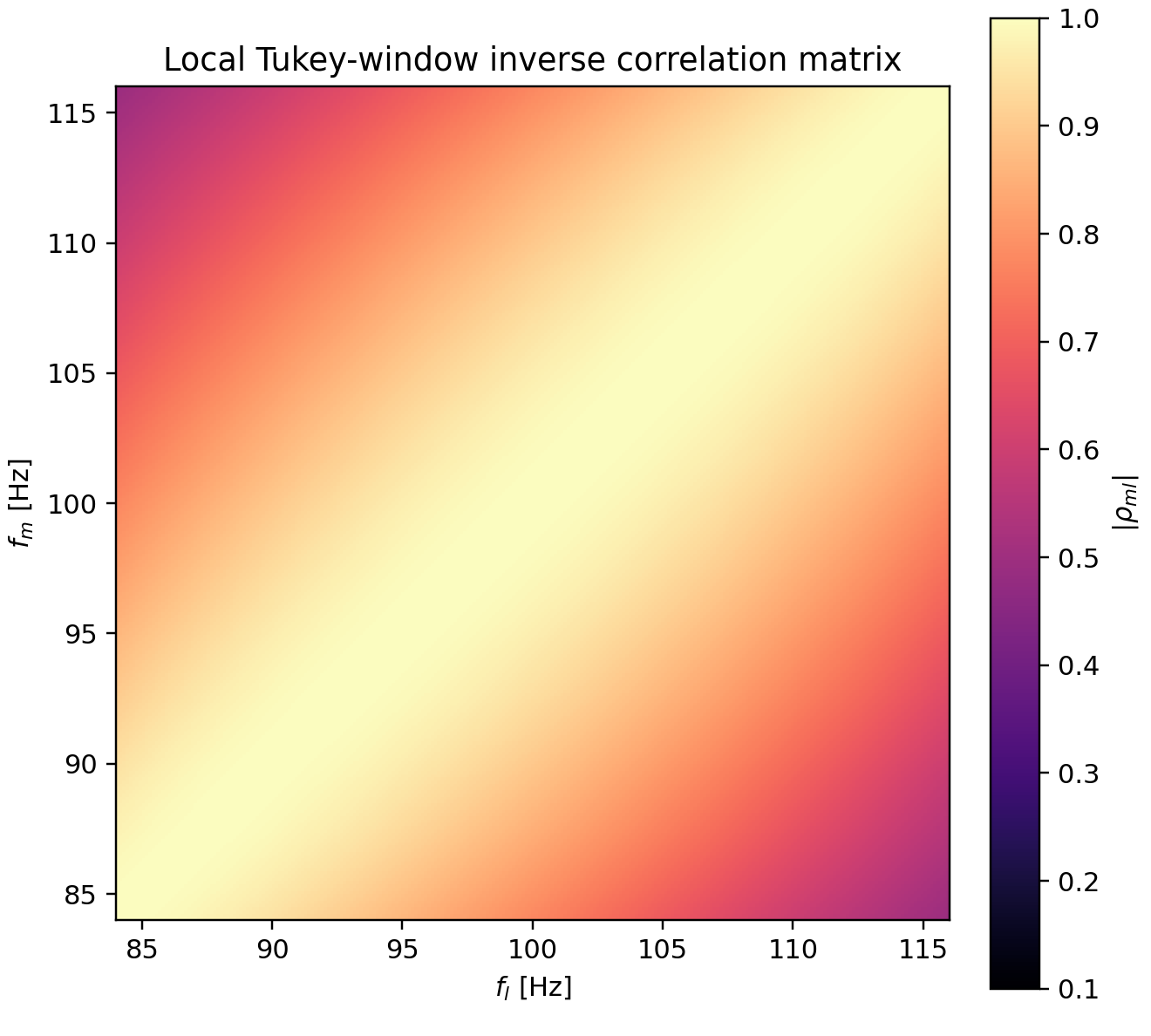} 
\caption{The noise correlation matrix and its inverse in a region near 100 Hz for otherwise stationary, flat spectrum Gaussian noise after a Tukey window has been applied. The inverse noise correlation matrix, which is used to compute the likelihood is dense, dramatically increasing the cost of the analyses relative to the stationary case. \label{fig:win}}
\end{figure}

The complex off-diagonal terms of the Fourier domain noise covariance matrix are what encode the
``when'' of a time-dependent noise level. Consider the likelihood contribution
from a signal $h$,
\begin{equation}
\rho^2 = h^\dagger C^{-1} h\, .
\end{equation}
If the covariance is approximated as diagonal in Fourier space, a time shift of
the signal,
\begin{equation}
\tilde h_m \rightarrow \tilde h_m e^{-2\pi i f_m t_0}\, ,
\end{equation}
does not change the signal-to-noise ratio, since only $|\tilde h_m|^2$ enters.
Thus a diagonal approximation can only represent a time-averaged noise level.
By contrast, with a full inverse covariance,
\begin{equation}
\rho^2 =
\sum_{ml} \tilde h_m^* (C^{-1})_{ml} \tilde h_l\, ,
\end{equation}
a time shift introduces phase factors
\begin{equation}
\tilde h_m^* \tilde h_l \rightarrow
\tilde h_m^* \tilde h_l e^{2\pi i (f_m-f_l)t_0}\, .
\end{equation}
These phases interfere with the complex phases of the off-diagonal elements of $C^{-1}$ and encode where the signal is suppressed or enhanced. As an explicit example, consider a flat noise spectrum, 8 seconds of data, and a Tukey window with 1 second roll-off. A plot of the noise correlation matrix and its inverse after the window has been applied is shown in Figure~\ref{fig:win}.

The inverse noise correlation matrix, which appears in the likelihood, is very dense. Computing likelihoods with this matrix would add considerable cost to an analysis.
 
It is interesting to compute the signal-to-noise (SNR) for a signal that is well-localized in time. For simplicity, consider a Gaussian windowed sinusoid with central frequency 100 Hz and one-sigma extent in time equal to 0.1 seconds. We can then compute the SNR as a function of the central time. We compute it three ways (1) using the full inverse covariance matrix (2) using the naive $C_{mm}^{-1}$ inverse and (3) the simple $C_{mm}^{-1}$ inverse with a correction which accounts for the loss of power due to the window function~\cite{Talbot:2025vth}.

 \begin{figure}[htp]
\includegraphics[width=0.45\textwidth]{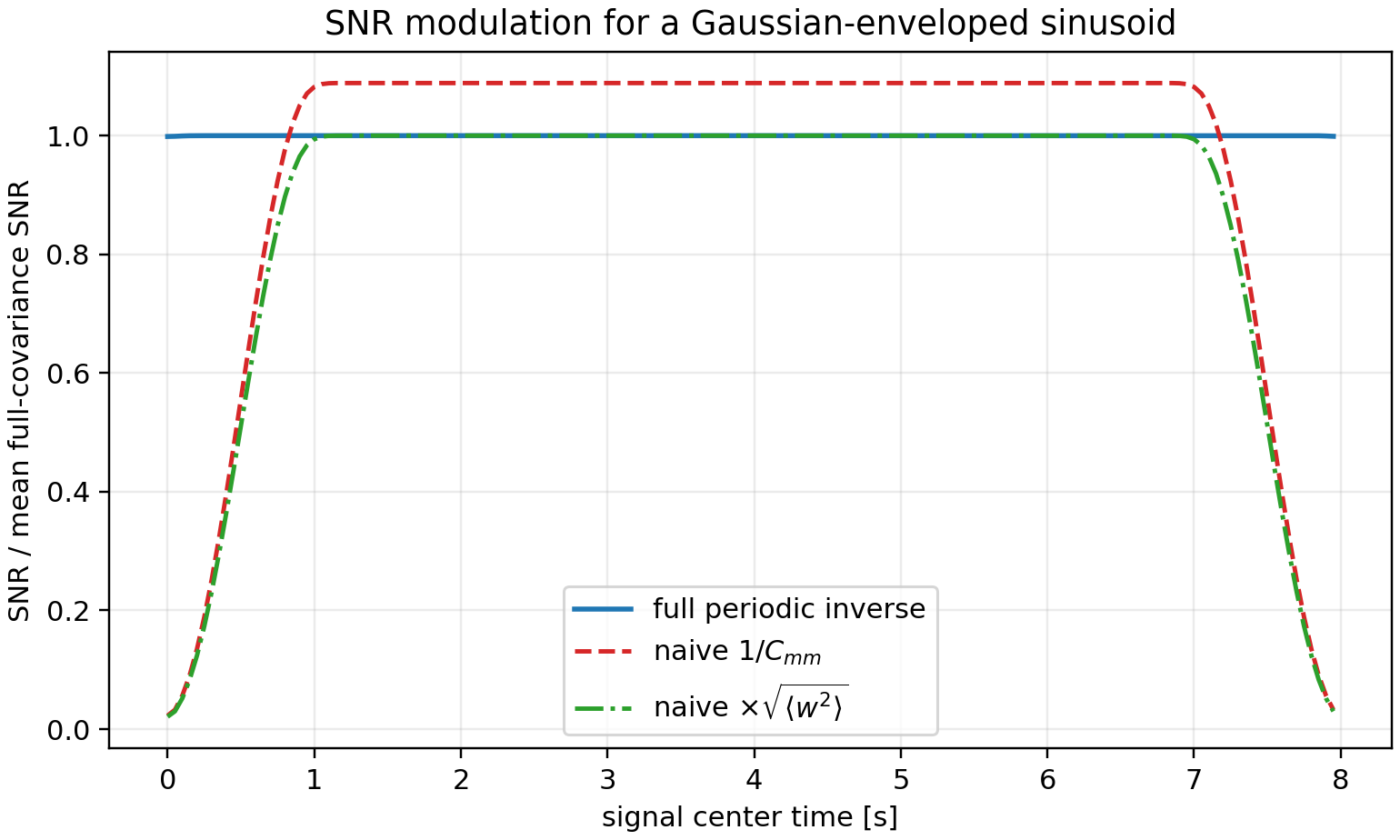} 
\caption{The SNR as a function of central time for a well-localized signal using the full inverse noise covariance matrix and the standard inverse of a diagonal covariance matrix, with and without the Tukey scale factor correction. The Tukey-corrected naive diagonal approximation comes very close to the full expression in regions away from the Tukey roll-off.}
\end{figure}

The SNR computed with the full noise covariance matrix remains flat across the 8-seconds. Both the signal and the noise are multiplied by the same window factor, and those factors cancel in the signal-to-noise ratio. In contrast, the naive diagonal approximation that is used in standard gravitational wave analyses shows a loss of SNR in the Tukey roll-off region. The SNR also overshoots in the region where the Tukey window is flat unless a correction factor is applied~\cite{Talbot:2025vth}. For signals that are finite in time and located in a region where the Tukey window is flat, it is possible to skip using the very expensive full inverse covariance matrix, and instead use the much faster diagonal approximation with the window correction factor. This approach does not work for long duration signals that run through the roll-off region.

\subsection{Finite Duration Effects}

The preceding discussion assumed that the data were finite and periodic. In reality, the data analyzed in gravitational wave astronomy are usually finite segments of much longer duration time series.
A slightly different description is needed if the data segment is a truncation of a much longer
stationary process with a continuum spectrum. In that case the finite-time
Fourier coefficient at bin $m$ is a projection of the continuum process through
the spectral response of the modulation/window. Let
\begin{equation}
\tilde{G}(\nu) = \sum_{n=0}^{N-1} g[n] e^{-2\pi i \nu t_n}\, ,
\end{equation}
where $g[n]$ includes any deterministic time-domain modulation and window
function. For a stationary process with continuum power spectrum $S(\nu)$, the
covariance between finite-time Fourier coefficients is
\begin{equation}
C_{ml} = \int d\nu\, S(\nu)\, \tilde{G}(f_m-\nu) \tilde{G}^*(f_l-\nu)\, .
\end{equation}
This expression makes explicit that the matrix elements are determined by the
overlap of two shifted spectral windows, weighted by the underlying spectrum.
For a rectangular window $g[n]=1$, $\tilde{G}(\nu)$ is the finite-duration cardinal sine
response. If $S(\nu)$ is exactly flat, the orthogonality of the Fourier basis
makes the covariance diagonal. If $S(\nu)$ varies across the width of the
spectral window, the cancellation is imperfect and off-diagonal correlations
appear. For a smooth spectrum these correlations scale roughly with the
fractional change of $S(\nu)$ over a bandwidth $1/T$. 

This effect is very important in pulsar timing array analyses~\cite{Allen:2024uqs,Crisostomi:2025vue} where the power spectra of the signals are very steep, and most of the information is contained in the first few discrete Fourier bins. These finite window correlations can usually be ignored for ground-based and space-based gravitational wave interferometer data as the noise rises sharply at low frequencies, and there is very little contribution to the likelihood from frequency bins where these correlations are significant.

\subsection{A Fourier domain model for non-stationary noise}

With these preliminaries out of the way, we are ready to define a model of non-stationary noise that proceeds by construction. The model is only approximate, and will not perfectly capture all non-stationary noise processes, but it is a very good model for the kinds of noise we encounter in gravitational wave astronomy.  
Suppose that the noise can be written as the sum of stationary noise processes $n_a(t)$ and a time variation $g_a(t)$:
\begin{equation}
x(t) = \sum_a n_a(t) g_a(t) \, .
\end{equation}
Further suppose that 
 the noise processes are stationary yet cross-correlated:
\begin{equation}
{\rm E}[{\tilde n}_a[j] {\tilde n}_b^*[k]] = N S_{ab}[k] \delta_{jk} 
\end{equation}
where the diagonal entries are the power spectra for the individual noise processes $S_{aa}[k] = S_{a}[k]$. The off-diagonal terms are in general complex, but the overall matrix $S_{ab}[k]$ is Hermitian, which is all we need for a valid covariance matrix. For this case we have
\begin{equation}
C_{ml} = \frac{1}{N} \sum_{ab} \sum_{k=0}^{N-1} S_{ab}[k] \tilde{g}_a[\{m-k\}_N]   \tilde{g}_b^*[\{l-k\}_N]\, .
\end{equation}
In the simpler case where the noise processes are uncorrelated we get
\begin{equation}\label{fcorr}
C_{ml} = \frac{1}{N} \sum_a \sum_{k=0}^{N-1} S_a[k] \tilde{g}_a[\{m-k\}_N]  \tilde{g}_a^*[\{l-k\}_N]\, ,
\end{equation}
where $S_a[k]$ is the power spectrum of the $n_a(t)$ process. This is the model we will adopt, following the argument given in the previous sub-section.

Now consider the case where we decompose the dynamic spectrum into independent frequency bins such that $\eta_k(t)$ represents a zero mean, unit variance stationary noise component associated with frequency bin $k$. Then we have
\begin{equation}
C_{ml} = \frac{1}{N}  \sum_{k=0}^{N-1} \tilde{s}_k[\{m-k\}_N ]   \tilde{s}_k^*[\{l-k\}_N]\, ,
\end{equation}
where $s^2_k[n] = S(f_k,t_n)$ is the discretely sampled dynamic spectrum and $\tilde{s}_k[m]$ is the discrete Fourier transform of $s_k[n]$. By construction, the spectrum is positive. The way to interpret this equation is that each contribution to the noise is made up of the product of a stationary noise component associated with a single frequency bin, and a time-dependent modulation function. The modulation then spreads the power over multiple frequency bins. This construction can be used to efficiently produce simulated non-stationary noise, as described in Appendix~\ref{sim}.

While this construction may appear fairly limited, it is actually quite general. Recall that any square integrable function $g(f,t)$, 
\begin{equation}
\int df\,dt\, |g(f,t)|^2 < \infty
\end{equation}
can be written as the infinite sum of separable functions:
\begin{equation}
g(f,t)=\sum_{k=0}^{\infty} a_k(f)b_k(t),
\end{equation}
which follows from a square integrable function $g(f,t)$ having a Schmidt expansion~\cite{Schmidt1907}
\begin{equation}
g(f,t)=\sum_{k=0}^{\infty} \sigma_k u_k(f)v_k(t).
\end{equation}
This is the continuous analog of the SVD of a matrix. While the sum is formally infinite, for ``well-behaved'', smoothly varying functions, the expansion is dominated by a modest number of eigenvalues, and the sum can be truncated without losing much accuracy. In our application, the time series is discretely sampled and finite, and $S(f,t)$ can always be written as a finite sum in that case.

\subsection{The wavelet domain noise covariance matrix}

As mentioned earlier, now that we have a model for non-stationary noise in the Fourier domain, we can map the model to any other representation using a linear transformation. It is possible to provide a simple explicit expression for the case of the Wilson-Daubechies (WD) family of wavelet transforms. The mapping from the frequency domain to the wavelet domain for WD wavelets can be computed using a discrete Fourier transform. Starting with frequency domain data $\tilde{x}[k]$, we compute the windowed Fourier transform
\begin{equation}
Z_i= \sum_{l = -N_t/2}^{N_t/2-1}
Q_i(l)  \tilde{x}\left[l+\frac{m_iN_t}{2}\right]\Phi[l] \, ,
\end{equation}
where
\begin{equation}
Q_i(l)=\exp\left(\frac{2\pi i(l+N_t/2)n_i}{N_t}\right)\, .
\end{equation}
For the interior positive-frequency layers, the time index $n_i$ runs from $0$ to $N_t-1$ and the frequency index $m_i$ runs from $1$ to $N_f-1$. The DC and Nyquist layers require modified pixels with half as many time samples; including those layers, the total number of real WD coefficients is $N_tN_f=N$. The function $\Phi[l]$ is the frequency domain representation of the window function. These windows must obey a partition of unity and orthogonality relation. The Meyer window function has seen extensive use in gravitational wave analyses, where it is referred to as the WDM transformation. Other window functions can also be used. The WD wavelet coefficients are then given by
\begin{equation}\label{fullf}
w_i =  \left\{\begin{array}{ll}
{\cal A}_f \Re Z_i, & (n_i+m_i)\;{\rm even} \\
(-1)^{m_i}{\cal A}_f \Im Z_i, & (n_i+m_i)\;{\rm odd}
\end{array}\right. \, .
\end{equation}
Equivalently,
\begin{equation}
w_i={\cal A}_f\left(\beta_i Z_i+\beta_i^* Z_i^*\right),
\end{equation}
where
\begin{equation}
\beta_i=\left\{\begin{array}{ll}
1/2, & (n_i+m_i)\;{\rm even}\\
(-1)^{m_i}/(2\I), & (n_i+m_i)\;{\rm odd}
\end{array}\right. .
\end{equation}
The normalization ${\cal A}_f$ is fixed by Parseval's theorem.  The complex windowed coefficients have covariance
\begin{equation}
K_{ij}={\rm E}[Z_iZ_j^*]
= \frac{1}{N}\sum_{k=0}^{N-1} U_{ki}U^*_{kj}
\end{equation}
and positive-negative covariance
\begin{equation}
H_{ij}={\rm E}[Z_iZ_j]
= \frac{1}{N}\sum_{k=0}^{N-1} U_{ki}V^*_{kj},
\end{equation}
where
\begin{eqnarray}
U_{ki} &=& \sum_l Q_i(l)\Phi[l]\,
\tilde{s}_k[\{a_i(l)-k\}_N] \, , \nonumber\\
V_{ki} &=& \sum_l Q_i^*(l)\Phi[l]\,
\tilde{s}_k[\{-a_i(l)-k\}_N] \, ,
\end{eqnarray}
and
\begin{equation}
a_i(l)=l+\frac{m_iN_t}{2} \, .
\end{equation}
The real wavelet domain noise covariance matrix $S_{ij} = {\rm E}[w_i w_j]$ is
\begin{equation}\label{wdmcov}
S_{ij} =
{\cal A}_f^2\left(
\beta_i\beta_j^*K_{ij}
+\beta_i^*\beta_j K_{ji}
+\beta_i\beta_jH_{ij}
+\beta_i^*\beta_j^*H_{ij}^*
\right) .
\end{equation}
Setting $S(f,t) = S(f)$ recovers the expression for the WD noise covariance matrix given in section III.A of Ref.~\cite{Cornish:2025awt}, while setting $S(f,t)= S(t)$ recovers the expressions given in section III.B of the same paper. The extension to the full Taylor series expansion of $S(f,t)$ is given in the appendix.

The formal expression (\ref{wdmcov}) can be very expensive to calculate directly. A much faster method, based on the Gram form of the covariance, is given in the appendix~\ref{fastcorr}.

\subsection{The Whittle Likelihood}

For completeness, we provide expressions for the Whittle likelihood for Gaussian noise. The usual expressions typically neglect the off-diagonal terms in the noise correlation matrix. In the Fourier domain we have
\begin{equation}
p(r)=\frac{1}{\pi^M \det C}
\exp\left(-r^\dagger C^{-1}r\right)
\end{equation}
where $r = d-h$ is the residual formed by subtracting the signal model $h$ from the data $d$. This expression skips the zero frequency and Nyquist layers which require special treatment as they are real. Thus $M=N/2-1$. In the wavelet domain we have
\begin{equation}
p(r)=\frac{1}{(2\pi)^{D/2} \sqrt{\det S}}
\exp\left(- \frac{1}{2}r_i S_{ij}^{-1}r_j\right)
\end{equation}
where $D$ is the number of real wavelet coefficients included. Again, the zero frequency and Nyquist layers require careful treatment as they have half as many time-frequency pixels as the central layers~\cite{Pearson:2025wfd}. 

\section{Relation to Wigner-Ville spectra}

The construction introduced here is closely related in spirit to the Wigner-Ville
approach.  The Wigner-Ville spectrum starts from a
two-time covariance and takes a Fourier transform in the time lag.  This gives a
high-resolution local spectral representation, but it is bilinear in the data and
can take negative values.  Various smoothed Wigner-Ville and related
constructions introduce additional averaging in time, frequency, or lag to reduce
these interference terms~\cite{lu2009wienerkhinchintheoremnonwidesense,Dechant_2015}.
Here we instead start with a positive dynamic spectrum, $S(f,t)=s^2(f,t)$, and
use it as an operational definition of the covariance matrix for finite data.

It is useful to see what Wigner-Ville construction would return for a noise process generated
from the amplitude functions $s(f,t)$.  To avoid distracting factors associated with
negative frequencies, consider the analytic, complex version of the construction,
\begin{equation}
z(t)=\int d\nu\, s(\nu,t)\,e^{2\pi i\nu t}\,dZ(\nu),
\end{equation}
where the random spectral increments obey
\begin{equation}
{\rm E}\left[dZ(\nu)dZ^*(\nu')\right]=\delta(\nu-\nu')\,d\nu d\nu' .
\end{equation}
Then the two-time covariance is
\begin{eqnarray}
&&{\rm E}\left[z(t+\tau/2)z^*(t-\tau/2)\right] \nonumber\\
&&\quad =\int d\nu\,
s(\nu,t+\tau/2)s^*(\nu,t-\tau/2)e^{2\pi i\nu\tau}.
\end{eqnarray}
Substituting this into the Wigner-Ville definition gives
\begin{eqnarray}
S_{\rm WV}(f,t)
&=& \int d\nu\int d\tau\,
e^{-2\pi i(f-\nu)\tau} \nonumber\\
&&\quad \times
s(\nu,t+\tau/2)s^*(\nu,t-\tau/2).
\end{eqnarray}
Thus the Wigner-Ville spectrum is not simply $s^2(f,t)$.  It is a lag transform
of the product of amplitude functions evaluated at the two times
$t+\tau/2$ and $t-\tau/2$.

In the slowly varying limit, and away from the boundaries of the data segment,
the two constructions agree.  For real positive $s$,
\begin{eqnarray}
&&s(\nu,t+\tau/2)s(\nu,t-\tau/2) \nonumber\\
&&\quad = S(\nu,t)
+\frac{\tau^2}{8}S(\nu,t)\partial_t^2\ln S(\nu,t)+\cdots .
\end{eqnarray}
Using this expansion in the previous expression gives
\begin{equation}
S_{\rm WV}(f,t)
\simeq S(f,t)
-\frac{1}{32\pi^2}
\partial_f^2\!\left[
S(f,t)\partial_t^2\ln S(f,t)
\right]+\cdots .
\end{equation}
There is no term proportional to the first time derivative because the
Wigner-Ville covariance uses the symmetric pair of times $t\pm\tau/2$.
Rapid time variation therefore appears as higher derivative corrections that
broaden or distort the local spectrum in frequency. The leading correction term can be positive of negative, and can drive the Wigner-Ville covariance negative. By construction, the dynamic spectrum $S(f,t)$ is always positive.

Our finite, discrete construction of the dynamic spectrum has an analogous relation to the Wigner-Ville form.
Writing the simulated time series schematically as
\begin{equation}
x[n]=\frac{1}{\sqrt{N}}\sum_{k=0}^{N-1}
s_k[n]\eta_k e^{2\pi i kn/N},
\end{equation}
with independent unit-variance $\eta_k$, the local lag covariance is
\begin{equation}
R[n,r]=\frac{1}{N}\sum_{k=0}^{N-1}
s_k[n+r/2]s_k^*[n-r/2]e^{2\pi i kr/N},
\end{equation}
up to bookkeeping for half-sample lags.  A discrete Wigner-Ville
transform would then give
\begin{equation}
W[p,n]=\frac{1}{N}\sum_{k=0}^{N-1}\sum_r
s_k[n+r/2]s_k^*[n-r/2]e^{-2\pi i(p-k)r/N}.
\end{equation}
If $s_k[n]$ is independent of $n$, this reduces to the stationary spectrum.
If $s_k[n]$ varies slowly over the lags that contribute appreciably, it
approximates $S(f_p,t_n)$.  The finite duration of the data introduces an
additional subtlety: if the DFT construction is interpreted periodically, the
lags wrap around the segment; if the data are zero outside the segment, the
allowed lag range depends on $n$ and introduces a time-dependent spectral
window.  In either case, the Wigner-Ville connection applies to the interior,
slowly varying approximation rather than an exact identity.

\section{Applications}

To illustrate the utility of the new non-stationary noise model, we provide two examples that are relevant to the future space-based gravitational wave detector LISA. The second example could equally apply to the LIGO instruments if we modified the frequency range and time scale. 

\begin{figure}[htp]
\includegraphics[width=0.45\textwidth]{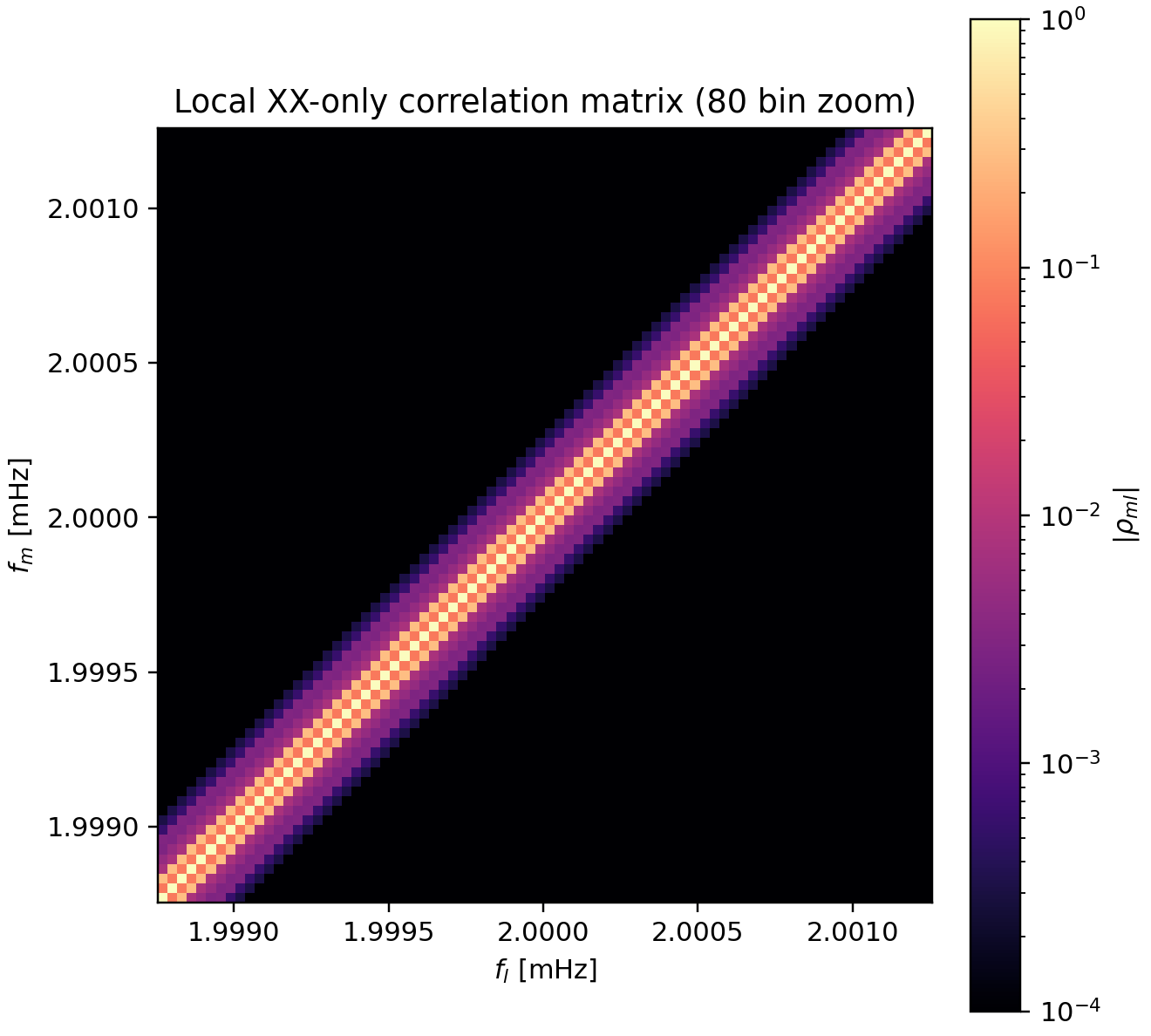} 
\includegraphics[width=0.45\textwidth]{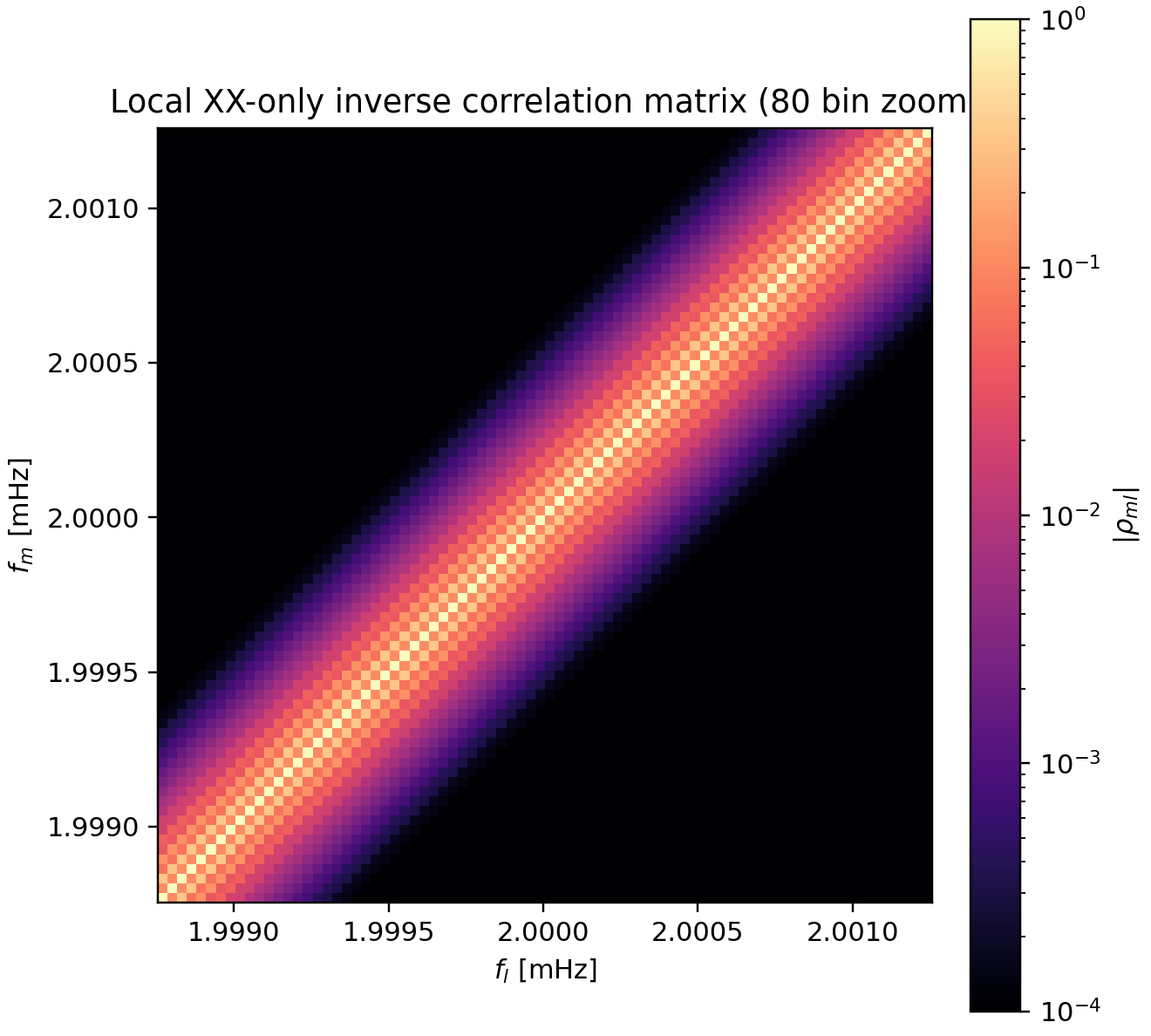} 
\caption{The galactic confusion noise correlation matrix and its inverse in a small band around 2 mHz for the TDI X-channel auto covariance.\label{fig:gal}}
\end{figure}

The first example is the modulation of the galactic confusion noise caused by the sweep of the antenna pattern as the LISA constellation orbits the Sun. The modulation has frequency components at harmonics of $1/{\rm yr}$, with the dominant terms coming from the $n=0,1,2$ harmonics~\cite{Digman:2022jmp}. This imparts side-bands on the Fourier domain noise covariance matrix. For simplicity, we compute the noise covariance matrix using exactly one year of data to avoid obscuring the modulation effects by window function effects. To keep the plots readable, we focus on a small frequency band around 2 mHz and only show the time-delay-interferometry (TDI) X-channel auto-covariance in Figure~\ref{fig:gal}.

\begin{figure}[htp]
\includegraphics[width=0.45\textwidth]{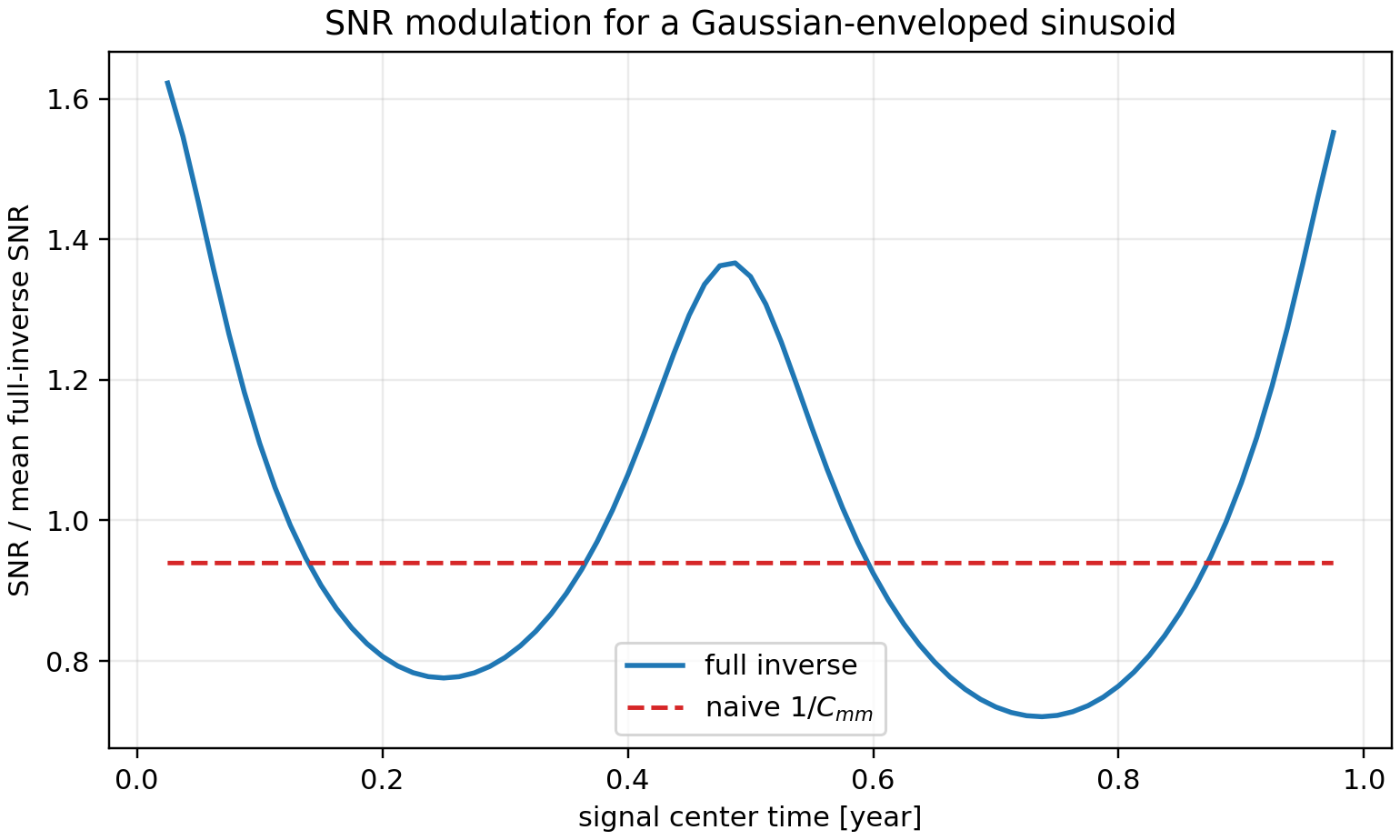}
\caption{The SNR as a function of central time for a well-localized signal using the full inverse noise covariance matrix and the standard inverse of a diagonal covariance matrix.\label{fig:galmod}}
\end{figure}

The additional significant side-bands in the inverse noise covariance matrix raise the computational cost of computing the frequency domain likelihood by a factor of 30 in this case. To illustrate the effect of the modulation on a short duration signal, Figure~\ref{fig:galmod} shows the SNR as a function of central time for a Gaussian windowed sinusoid with one-sigma extent of 1-day, and central frequency 2 mHz computed using the full inverse of the confusion noise covariance matrix and the diagonal approximation.

The time-varying confusion noise level causes the SNR to vary by a factor of two over the year. The diagonal approximation, which has been used for many years in studies of the LISA mission, completely misses this large effect. The SNR modulation was previously demonstrated using the WDM noise covariance matrix~\cite{Digman:2022jmp}.

\begin{figure}[htp]
\includegraphics[width=0.42\textwidth]{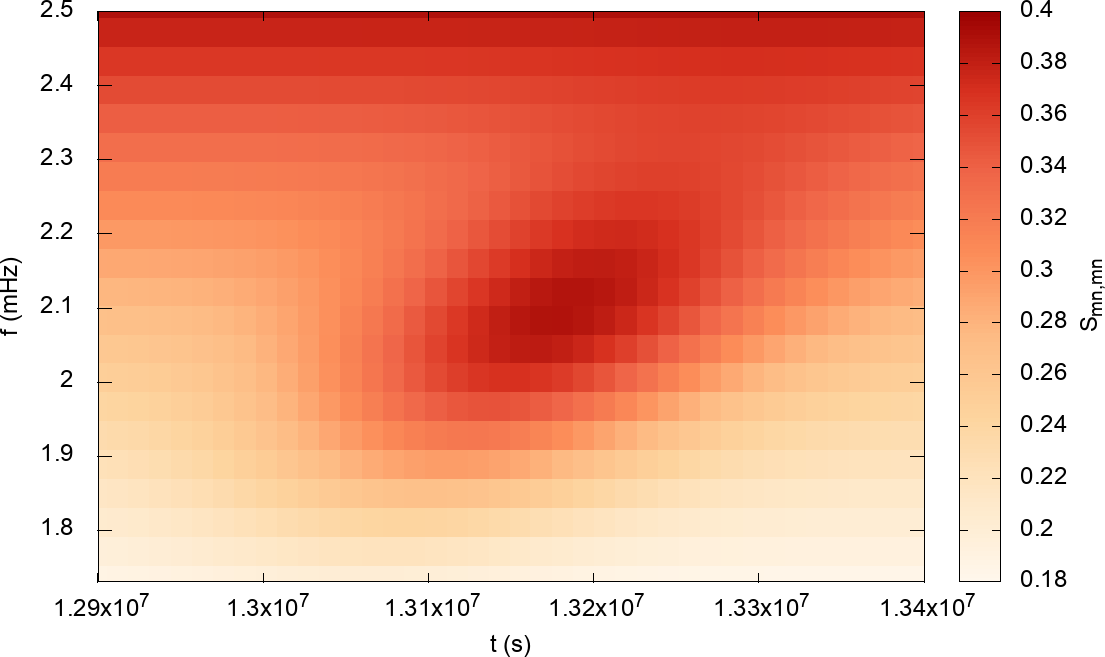} 
\caption{The diagonal entries of the WDM noise correlation matrix for the dynamic spectrum with a Gaussian bump. The tilted localized structure of the bump is clearly visible.\label{fig:wdmbump}}
\end{figure}

The previous examples were for cases where the dynamic spectrum $S(f,t)$ naturally factored into a product of a function of time and a function of frequency. For the final example we consider a case where the spectrum does not factor. The dynamic spectrum is described by an overall power law with a Gaussian bump in time and frequency with axes that are not aligned with the time and frequency directions:
\begin{equation}\label{bump}
S(f,t)=S_0\left(\frac{|f|}{f_{\rm ref}}\right)^\alpha \left[1+A\exp\left(-\frac{1}{2}Q(f,t)\right)\right],
\end{equation}
with quadratic form
\begin{equation}
Q(f,t)=\frac{x^2-2q xy+y^2}{1-q^2},
\end{equation}
where
\begin{equation}
x=\frac{|f|-f_0}{\sigma_f},
\qquad
y=\frac{t-t_0}{\sigma_t}.
\end{equation}
Figure~\ref{fig:wdmbump} shows the diagonal entries of the WDM noise correlation matrix. In this simulation the Gaussian bump had an amplitude of $A=0.5$, $t_0=5$ months, $f_0= 2$ mHz, $\sigma_t=8 \Delta T$, $\sigma_f = 4 \Delta F$, correlation $q =0.5$ and power law index $\alpha = 2$. Here $\Delta T$ and $\Delta F$ are the extents of the WDM pixels in time and frequency.

\begin{figure}[htp]
\includegraphics[width=0.45\textwidth]{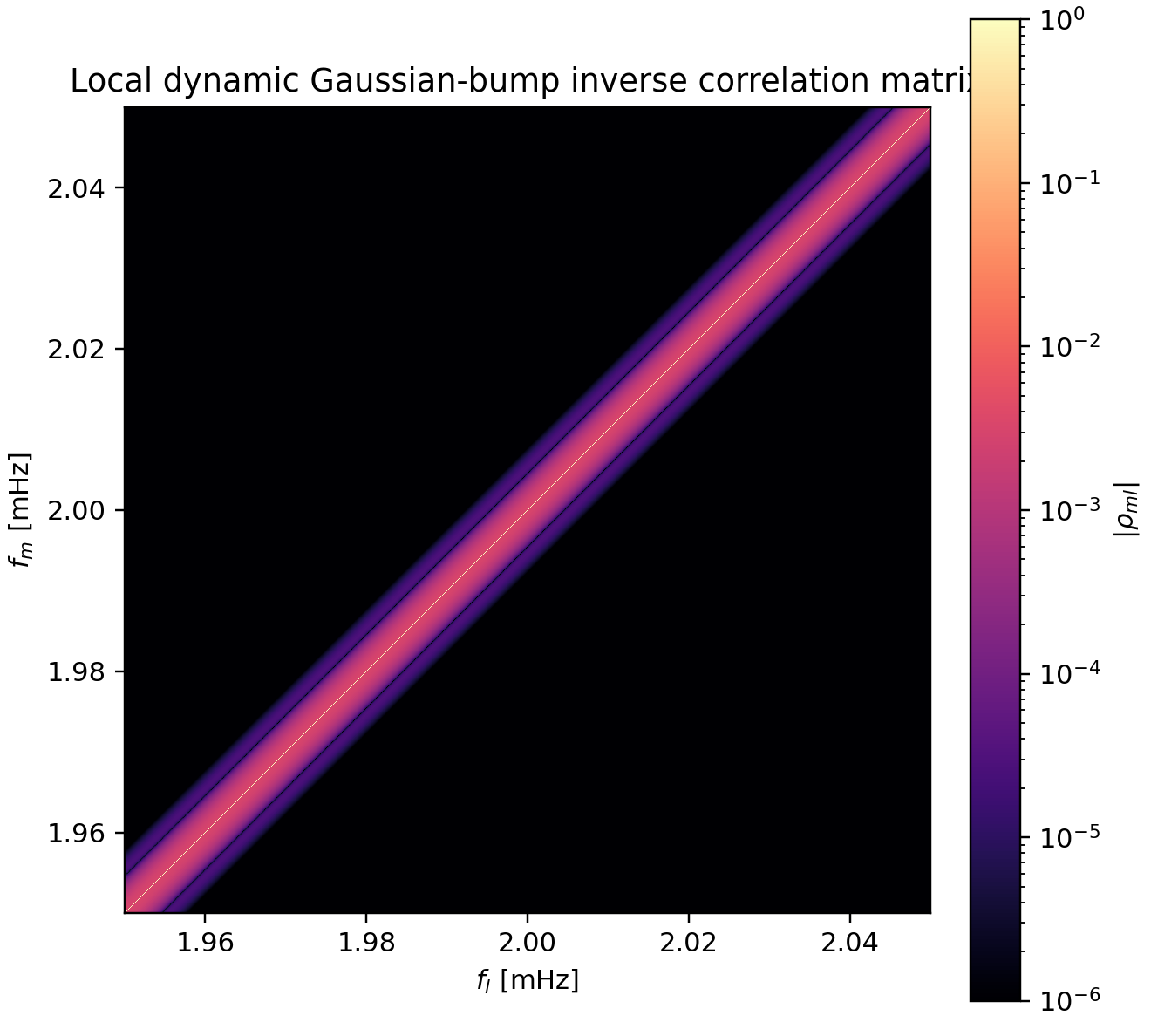} 
\caption{The inverse of the noise correlation matrix for a dynamic spectrum with a Gaussian bump that does not factor.\label{fig:bump}}
\end{figure}

Figure~\ref{fig:bump} shows the inverse of the Fourier domain noise covariance matrix for the dynamic spectrum (\ref{bump}). The noise covariance matrix and its inverse are very close to diagonal in this case, but it would be a mistake to drop these seemingly small terms. The small size of the off-diagonal entries of the Fourier domain noise covariance matrix follows from the Gaussian bump being well localized in time. Its influence gets spread and diluted in the frequency domain, but ignoring these small entries would remove all knowledge of the non-stationarity. 

The WDM noise covariance matrix also has off-diagonal terms, but these are mostly due to the overall $f^2$ power law, and not the Gaussian bump. The WDM pixel correlations can be rendered negligible by partially pre-whitening the data using the time average of the dynamic spectrum $S(f,t)$. This is an inexpensive step for an actual analysis as the data only has to be pre-whitened once, and the templates can be pre-whitened during the WDM transform since this is done using the Fourier to WDM mapping. Even if the model for the dynamic spectrum is updated during the analysis there is no need to change the pre-whitening as it does not need to be perfect. All we need for it to do is reduce the gradients across the pixels to an acceptable level. 

\begin{figure}[htp]
\includegraphics[width=0.45\textwidth]{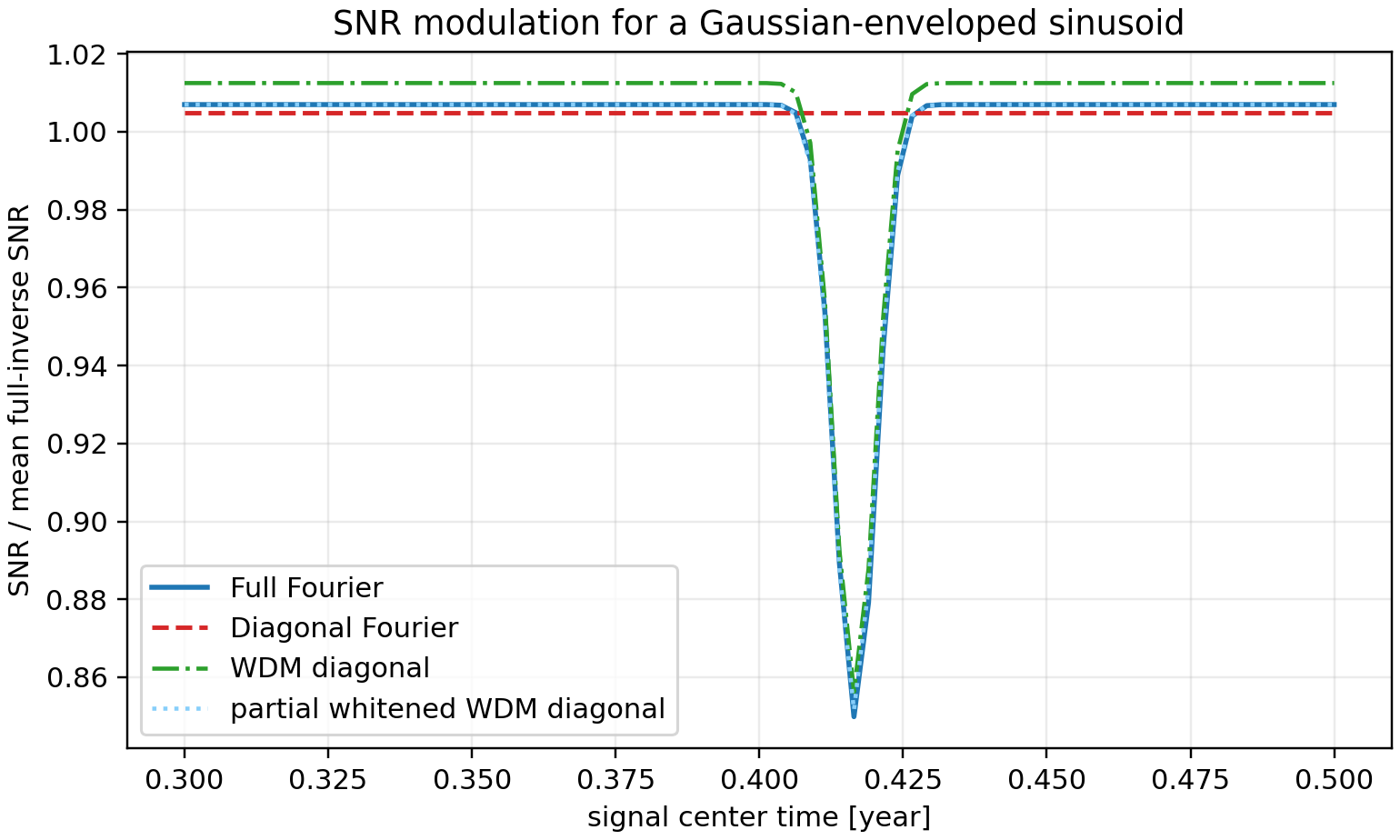} 
\caption{The SNR as a function of central time for a Gaussian enveloped sinusoid. The SNR is computed four different ways: using the full Fourier domain covariance matrix; the naive diagonal Fourier domain covariance matrix; the naive diagonal WDM covariance matrix; and the diagonal partially pre-whitened WDM noise covariance matrix.\label{fig:snrbump}}
\end{figure}

Figure~\ref{fig:snrbump} shows the SNR as a function of central time for the same Gaussian enveloped sinusoid that was used to study the impact of the modulated galactic confusion noise. Once again, we see that the diagonal approximation in the Fourier domain fails to capture the loss of SNR that occurs when the noise is elevated. The diagonal approximation to the WDM noise covariance matrix does a much better job, but it over-estimates the SNR by $\sim 0.5\%$. This is due to off-diagonal terms in the covariance matrix that mostly come from the power law portion of the dynamic spectrum, and not the bump in the power spectrum. Including the dominant $\Delta m = 0$, $\Delta n = \pm 1$ off-diagonal terms in the WDM covariance matrix is enough to reduce the error to $\sim 0.1\%$. But a simpler, and more accurate approach is to partially pre-whiten the Fourier domain data and template using the time average of the spectrum $\bar{S}(f) =\frac{1}{T} \int_0^T S(f,t) dt$ prior to applying the WDM transform, which reduces the error to $\sim 0.002\%$. Of course, all of these numbers are tiny compared to the $\sim 15\%$ error of the diagonal Fourier domain approximation.

\section{Discussion}

We have introduced a Gram-factored model for non-stationary noise that is well behaved and easy to calculate. This model can serve as the foundation for analyses of gravitational wave data that account for non-stationary noise. Incorporating non-stationarity will become increasingly important as the detector sensitivities improve and signals spend long times in the sensitive band of next generation gravitational wave detectors. In a follow up study~\cite{ghoshcornish2026}, we introduce a method to estimate $S(f,t)$ from detector data, and study the impact of non-stationarity on the analysis of low mass compact binary mergers.
 
 \section*{Acknowledgments}
This work was supported by the Simons Foundation award SFI-MPS-BH-00012593-04, the NASA LISA Preparatory Science Grant 80NSSC19K0320, and NSF awards PHY2513363.

\appendix

\section{Focused wavelet-domain covariance calculations\label{fastcorr}}

One way to compute a wavelet-domain covariance is to first build a dense block
of the Fourier domain covariance $C_{ab}$ for all Fourier bins touched by the wavelet windows, and then
contract this block with the wavelet filters.  This is wasteful if only a small
neighborhood of WDM pixels is needed.  A faster approach is to push the WDM
projection inside the sum over $k$ in (\ref{main}).

Let $i$ denote a selected WDM pixel $(m_i,n_i)$, and let
\begin{equation}
a_i(l)=l+\frac{m_iN_t}{2}, \qquad
Q_i(l)=\exp\left(\frac{2\pi i(l+N_t/2)n_i}{N_t}\right) .
\end{equation}
For each basis frequency $k$, define the projected response
\begin{equation}
U_{ki} =
\sum_l Q_i(l)\Phi[l]\,
\tilde{s}_k[\{a_i(l)-k\}_N] .
\end{equation}
Then the covariance of the complex windowed WDM coefficients is the small Gram
matrix
\begin{equation}\label{Kfactor}
K_{ij} = {\rm E}[Z_i Z_j^*]
= \frac{1}{N}\sum_{k=0}^{N-1} U_{ki} U_{kj}^* .
\end{equation}
The real WDM coefficients follow from the same parity projection used in the
transform,
\begin{equation}
w_i={\cal A}_f\left(\beta_i Z_i+\beta_i^* Z_i^*\right),
\end{equation}
where $\beta_i=1/2$ for even $n_i+m_i$ and
$\beta_i=(-1)^{m_i}/(2\I)$ for odd $n_i+m_i$. The positive-negative frequency contribution can be computed by defining
\begin{equation}
V_{ki} =
\sum_l Q_i^*(l)\Phi[l]\,
\tilde{s}_k[\{-a_i(l)-k\}_N] ,
\end{equation}
which gives
\begin{equation}
H_{ij} = {\rm E}[Z_iZ_j]
= \frac{1}{N}\sum_{k=0}^{N-1} U_{ki}V_{kj}^* .
\end{equation}
The exact real WDM covariance is then
\begin{equation}
S_{ij} =
{\cal A}_f^2\left(
\beta_i\beta_j^*K_{ij}
+\beta_i^*\beta_j K_{ji}
+\beta_i\beta_jH_{ij}
+\beta_i^*\beta_j^*H_{ij}^*
\right) .
\end{equation}
For interior positive-frequency layers, the $H_{ij}$ term is usually separated
from the $K_{ij}$ term by a large harmonic offset and is strongly suppressed.
Near the zero frequency and Nyquist layers, or for dynamic spectra with broad harmonic
content, the $H_{ij}$ term should be retained.

The practical speed-up comes from two restrictions.  First, the set of pixels
$i$ is small; for example, the local correlations span at most $3 \times 17$ pixels for the WDM
transform used in this paper, and of those, just the central few terms are significant.
Second, the sum over $k$ does not need all basis frequencies.
Only values of $k$ for which the harmonics of
$\tilde{s}_k[\{a_i-k\}_N]$ overlap the WDM supports contribute
appreciably.  For slowly varying, or localized dynamic spectra these harmonics
are compact, so one can retain a finite band ${\cal K}$ around the selected
supports and check convergence by increasing its width:
\begin{equation}
K_{ij} \simeq \frac{1}{N}\sum_{k\in{\cal K}} U_{ki}U_{kj}^* .
\end{equation}
When the $H_{ij}$ term is retained, the analogous restricted sum is used for
that term as well, with the retained band enlarged until the local covariance
has converged.
This avoids forming any dense Fourier covariance block.  The computational cost
scales with the number of retained basis bins, the number of selected WDM
pixels, and the width of the WDM frequency windows.

If the data have been partially whitened in the Fourier domain by a stationary
reference spectrum $\bar{S}(f)$, the same operation is included by inserting
the diagonal weight $\bar{S}^{-1/2}$ in the WDM projection:
\begin{equation}
U_{ki}^{\rm white} =
\sum_l Q_i(l)\Phi[l]\,
\frac{\tilde{s}_k[\{a_i(l)-k\}_N]}
{\sqrt{\bar{S}(f_{a_i(l)})}} .
\end{equation}
The rest of the calculation is unchanged.

\subsection{Taylor expansion of the WDM noise covariance matrix\label{taylor}}

In Ref.~\cite{Cornish:2025awt} the WDM noise covariance matrix was calculated analytically for the special cases $S(f,t) = S_f(f)$ and $S(f,t) = S_t(t)$. Now with the full expression for the WD noise covariance in hand we are in a position to complete the calculation to include cross terms in the Taylor expansion of $S(f,t)$ about a WD pixel center.

The mixed corrections can be organized by expanding the dynamic spectrum about
the center of a WDM pixel,
\begin{equation}
S(f,t)=S_0\sum_{p,q}
\frac{c_{pq}}{p!\,q!}\delta f^p\delta t^q ,
\end{equation}
where $\delta f=(f-f_m)/\Delta F$,
$\delta t=(t-t_n)/\Delta T$, and
\begin{equation}
c_{pq} =
\frac{\Delta F^p\Delta T^q}{S_0}
\frac{\partial^{p+q}S}{\partial f^p\partial t^q}\, .
\end{equation}
The pure-frequency coefficients $c_{p0}$ and pure-time coefficients $c_{0q}$
reduce to the quantities $s_p$ and $\mu_q$ used in Ref.~\cite{Cornish:2025awt}.  The mixed coefficients
can be evaluated numerically by inserting the expansion
$S(f,t)=S_0(1+\epsilon\delta f^p\delta t^q)$ into Eq.~(\ref{Kfactor}) and
taking the first derivative with respect to $\epsilon$.  This provides a useful
check on the signs: the $c_{10}$, $c_{20}$, $c_{01}$ and $c_{02}$ coefficients
reproduce Eqs.~(35) and (43) of Ref.~\cite{Cornish:2025awt}.

For the same Meyer-window choice used in Ref.~\cite{Cornish:2025awt}, $d=6$, $A=0$ and
$B=\Delta\Omega$, the leading mixed terms are
\begin{eqnarray}
&& S_{mn,m(n\pm 1)} =
(-1)^{n+1}X\,0.105\,c_{11}+\dots \nonumber\\
&& S_{mn,(m\pm1)n} =
(-1)^{n+m}X\,0.125\,c_{11}+\dots \nonumber\\
&& S_{mn,(m\pm1)(n\pm1)} =
(-1)^m\epsilon\tau X\,0.111\,c_{11}+\dots \nonumber\\
&& S_{mn,(m\pm1)(n\pm2)} =
(-1)^{n+m+1}X\,0.0748\,c_{11}+\dots ,
\end{eqnarray}
where $\epsilon=+1$ for the upper frequency neighbor $m+1$ and
$\epsilon=-1$ for $m-1$, while $\tau=+1$ for $n+r$ and $\tau=-1$ for $n-r$.
Here $X$ denotes the leading flat-spectrum diagonal power in the reference WDM
pixel. For a separable spectrum $S(f,t)=S_f(f)S_t(t)$, $c_{11}=c_{10}c_{01}$, so the
mixed term is second order in the local fractional gradients.  A genuinely
tilted or non-separable dynamic spectrum also contributes through
$\Delta F\Delta T\,\partial_f\partial_t\ln S$.

\subsection{Generating noise realizations\label{sim}}

The same factorization used to compute the WD noise correlation matrix efficiently gives a direct way to simulate frequency-domain data
with dynamic spectrum $S(f,t)$.  The usual way to generate Gaussian data with a
specified covariance matrix is to find a matrix square root, for example by a
Cholesky decomposition or by diagonalizing the covariance matrix and taking the
square root of its eigenvalues.  Here that extra step is unnecessary: the Gram
form of the covariance already provides a square-root factor.  Let $\eta_k$ be
independent proper complex unit-variance Gaussian variates in the basis-frequency index,
with ${\rm E}[\eta_k\eta_l^*]=\delta_{kl}$ and ${\rm E}[\eta_k\eta_l]=0$.
Define
\begin{equation}
\tilde{x}[a] =
\frac{1}{\sqrt{N}}
\sum_{k=0}^{N-1}
\tilde{s}_k[\{a-k\}_N]\,\eta_k .
\end{equation}
Then
\begin{eqnarray}
{\rm E}\left[\tilde{x}[a]\tilde{x}^*[b]\right]
&=& \frac{1}{N}
\sum_{k,l}
\tilde{s}_k[\{a-k\}_N]
\tilde{s}_l^*[\{b-l\}_N]
{\rm E}[\eta_k\eta_l^*] \nonumber \\
&=& \frac{1}{N}
\sum_{k=0}^{N-1}
\tilde{s}_k[\{a-k\}_N]
\tilde{s}_k^*[\{b-k\}_N] \, ,
\end{eqnarray}
which is the desired covariance matrix.  In matrix notation this is simply
$\tilde{x}=A\eta$, with
\begin{equation}
A_{ak} = \frac{1}{\sqrt{N}}\,
\tilde{s}_k[\{a-k\}_N] .
\end{equation}
Thus the simulation method and the covariance calculation are based on the
same square-root factor of the Fourier-domain covariance.

For a real time-domain noise realization one imposes the usual Hermitian
symmetry on the Fourier coefficients, or equivalently, works with an
independent real basis such as cosine and sine components.  In the special
separable case
\begin{equation}
S(f,t)=S_f(f)J(t),
\end{equation}
the simulation can be performed more simply.  Draw stationary colored
Fourier-domain noise $\tilde{y}[k]$ with variance proportional to $S_f(f_k)$,
transform it to the time domain, multiply by $\sqrt{J(t)}$, and transform back:
\begin{equation}
\tilde{x} =
{\cal F}
\left[\sqrt{J(t)}\,{\cal F}^{-1}[\tilde{y}]\right] .
\end{equation}
This separable construction is just a fast implementation of the same
$A_{ak}$ covariance square root when the time dependence is common to all
basis frequencies.

\bibliography{refs}

\begin{thebibliography}{28}
\expandafter\ifx\csname natexlab\endcsname\relax\def\natexlab#1{#1}\fi
\expandafter\ifx\csname bibnamefont\endcsname\relax
  \def\bibnamefont#1{#1}\fi
\expandafter\ifx\csname bibfnamefont\endcsname\relax
  \def\bibfnamefont#1{#1}\fi
\expandafter\ifx\csname citenamefont\endcsname\relax
  \def\citenamefont#1{#1}\fi
\expandafter\ifx\csname url\endcsname\relax
  \def\url#1{\texttt{#1}}\fi
\expandafter\ifx\csname urlprefix\endcsname\relax\def\urlprefix{URL }\fi
\providecommand{\bibinfo}[2]{#2}
\providecommand{\eprint}[2][]{\url{#2}}

\bibitem[{\citenamefont{Finn}(1992)}]{Finn:1992wt}
\bibinfo{author}{\bibfnamefont{L.~S.} \bibnamefont{Finn}},
  \bibinfo{journal}{Phys. Rev. D} \textbf{\bibinfo{volume}{46}},
  \bibinfo{pages}{5236} (\bibinfo{year}{1992}), \eprint{gr-qc/9209010}.

\bibitem[{\citenamefont{Abbott et~al.}(2020)}]{LIGOScientific:2019hgc}
\bibinfo{author}{\bibfnamefont{B.~P.} \bibnamefont{Abbott}}
  \bibnamefont{et~al.} (\bibinfo{collaboration}{LIGO Scientific, Virgo}),
  \bibinfo{journal}{Class. Quant. Grav.} \textbf{\bibinfo{volume}{37}},
  \bibinfo{pages}{055002} (\bibinfo{year}{2020}), \eprint{1908.11170}.

\bibitem[{\citenamefont{Cornish and Littenberg}(2015)}]{Cornish:2014kda}
\bibinfo{author}{\bibfnamefont{N.~J.} \bibnamefont{Cornish}} \bibnamefont{and}
  \bibinfo{author}{\bibfnamefont{T.~B.} \bibnamefont{Littenberg}},
  \bibinfo{journal}{Class. Quant. Grav.} \textbf{\bibinfo{volume}{32}},
  \bibinfo{pages}{135012} (\bibinfo{year}{2015}), \eprint{1410.3835}.

\bibitem[{\citenamefont{Cornish et~al.}(2020)\citenamefont{Cornish, Littenberg,
  B\'ecsy, Chatziioannou, Clark, Ghonge, and Millhouse}}]{Cornish:2020dwh}
\bibinfo{author}{\bibfnamefont{N.~J.} \bibnamefont{Cornish}},
  \bibinfo{author}{\bibfnamefont{T.~B.} \bibnamefont{Littenberg}},
  \bibinfo{author}{\bibfnamefont{B.}~\bibnamefont{B\'ecsy}},
  \bibinfo{author}{\bibfnamefont{K.}~\bibnamefont{Chatziioannou}},
  \bibinfo{author}{\bibfnamefont{J.~A.} \bibnamefont{Clark}},
  \bibinfo{author}{\bibfnamefont{S.}~\bibnamefont{Ghonge}}, \bibnamefont{and}
  \bibinfo{author}{\bibfnamefont{M.}~\bibnamefont{Millhouse}}
  (\bibinfo{year}{2020}), \eprint{2011.09494}.

\bibitem[{\citenamefont{Chatziioannou et~al.}(2021)\citenamefont{Chatziioannou,
  Cornish, Wijngaarden, and Littenberg}}]{Chatziioannou:2021ezd}
\bibinfo{author}{\bibfnamefont{K.}~\bibnamefont{Chatziioannou}},
  \bibinfo{author}{\bibfnamefont{N.}~\bibnamefont{Cornish}},
  \bibinfo{author}{\bibfnamefont{M.}~\bibnamefont{Wijngaarden}},
  \bibnamefont{and} \bibinfo{author}{\bibfnamefont{T.~B.}
  \bibnamefont{Littenberg}}, \bibinfo{journal}{Phys. Rev. D}
  \textbf{\bibinfo{volume}{103}}, \bibinfo{pages}{044013}
  (\bibinfo{year}{2021}), \eprint{2101.01200}.

\bibitem[{\citenamefont{Hourihane et~al.}(2022)\citenamefont{Hourihane,
  Chatziioannou, Wijngaarden, Davis, Littenberg, and
  Cornish}}]{Hourihane:2022doe}
\bibinfo{author}{\bibfnamefont{S.}~\bibnamefont{Hourihane}},
  \bibinfo{author}{\bibfnamefont{K.}~\bibnamefont{Chatziioannou}},
  \bibinfo{author}{\bibfnamefont{M.}~\bibnamefont{Wijngaarden}},
  \bibinfo{author}{\bibfnamefont{D.}~\bibnamefont{Davis}},
  \bibinfo{author}{\bibfnamefont{T.}~\bibnamefont{Littenberg}},
  \bibnamefont{and} \bibinfo{author}{\bibfnamefont{N.}~\bibnamefont{Cornish}},
  \bibinfo{journal}{Phys. Rev. D} \textbf{\bibinfo{volume}{106}},
  \bibinfo{pages}{042006} (\bibinfo{year}{2022}), \eprint{2205.13580}.

\bibitem[{\citenamefont{Zackay et~al.}(2019)\citenamefont{Zackay, Venumadhav,
  Roulet, Dai, and Zaldarriaga}}]{Zackay:2019kkv}
\bibinfo{author}{\bibfnamefont{B.}~\bibnamefont{Zackay}},
  \bibinfo{author}{\bibfnamefont{T.}~\bibnamefont{Venumadhav}},
  \bibinfo{author}{\bibfnamefont{J.}~\bibnamefont{Roulet}},
  \bibinfo{author}{\bibfnamefont{L.}~\bibnamefont{Dai}}, \bibnamefont{and}
  \bibinfo{author}{\bibfnamefont{M.}~\bibnamefont{Zaldarriaga}}
  (\bibinfo{year}{2019}), \eprint{1908.05644}.

\bibitem[{\citenamefont{Mozzon et~al.}(2020)\citenamefont{Mozzon, Nuttall,
  Lundgren, Dent, Kumar, and Nitz}}]{Mozzon_2020}
\bibinfo{author}{\bibfnamefont{S.}~\bibnamefont{Mozzon}},
  \bibinfo{author}{\bibfnamefont{L.~K.} \bibnamefont{Nuttall}},
  \bibinfo{author}{\bibfnamefont{A.}~\bibnamefont{Lundgren}},
  \bibinfo{author}{\bibfnamefont{T.}~\bibnamefont{Dent}},
  \bibinfo{author}{\bibfnamefont{S.}~\bibnamefont{Kumar}}, \bibnamefont{and}
  \bibinfo{author}{\bibfnamefont{A.~H.} \bibnamefont{Nitz}},
  \bibinfo{journal}{Classical and Quantum Gravity}
  \textbf{\bibinfo{volume}{37}}, \bibinfo{pages}{215014}
  (\bibinfo{year}{2020}), ISSN \bibinfo{issn}{1361-6382},
  \urlprefix\url{http://dx.doi.org/10.1088/1361-6382/abac6c}.

\bibitem[{\citenamefont{Digman and Cornish}(2022)}]{Digman:2022jmp}
\bibinfo{author}{\bibfnamefont{M.~C.} \bibnamefont{Digman}} \bibnamefont{and}
  \bibinfo{author}{\bibfnamefont{N.~J.} \bibnamefont{Cornish}},
  \bibinfo{journal}{Astrophys. J.} \textbf{\bibinfo{volume}{940}},
  \bibinfo{pages}{10} (\bibinfo{year}{2022}), \eprint{2206.14813}.

\bibitem[{\citenamefont{Mozzon et~al.}(2022)\citenamefont{Mozzon, Ashton,
  Nuttall, and Williamson}}]{Mozzon:2021wam}
\bibinfo{author}{\bibfnamefont{S.}~\bibnamefont{Mozzon}},
  \bibinfo{author}{\bibfnamefont{G.}~\bibnamefont{Ashton}},
  \bibinfo{author}{\bibfnamefont{L.~K.} \bibnamefont{Nuttall}},
  \bibnamefont{and} \bibinfo{author}{\bibfnamefont{A.~R.}
  \bibnamefont{Williamson}}, \bibinfo{journal}{Phys. Rev. D}
  \textbf{\bibinfo{volume}{106}}, \bibinfo{pages}{043504}
  (\bibinfo{year}{2022}), \eprint{2110.11731}.

\bibitem[{\citenamefont{Allen and Romano}(2025)}]{Allen:2024uqs}
\bibinfo{author}{\bibfnamefont{B.}~\bibnamefont{Allen}} \bibnamefont{and}
  \bibinfo{author}{\bibfnamefont{J.~D.} \bibnamefont{Romano}},
  \bibinfo{journal}{Phys. Rev. Lett.} \textbf{\bibinfo{volume}{134}},
  \bibinfo{pages}{031401} (\bibinfo{year}{2025}), \eprint{2407.10968}.

\bibitem[{\citenamefont{Crisostomi et~al.}(2025)\citenamefont{Crisostomi, van
  Haasteren, Meyers, and Vallisneri}}]{Crisostomi:2025vue}
\bibinfo{author}{\bibfnamefont{M.}~\bibnamefont{Crisostomi}},
  \bibinfo{author}{\bibfnamefont{R.}~\bibnamefont{van Haasteren}},
  \bibinfo{author}{\bibfnamefont{P.~M.} \bibnamefont{Meyers}},
  \bibnamefont{and}
  \bibinfo{author}{\bibfnamefont{M.}~\bibnamefont{Vallisneri}}
  (\bibinfo{year}{2025}), \eprint{2506.13866}.

\bibitem[{\citenamefont{Ghosh and Cornish}(2026)}]{ghoshcornish2026}
\bibinfo{author}{\bibfnamefont{S.}~\bibnamefont{Ghosh}} \bibnamefont{and}
  \bibinfo{author}{\bibfnamefont{N.~J.} \bibnamefont{Cornish}},
  \bibinfo{journal}{In Preparation}  (\bibinfo{year}{2026}).

\bibitem[{\citenamefont{Wigner}(1932)}]{Wigner:1932}
\bibinfo{author}{\bibfnamefont{E.~P.} \bibnamefont{Wigner}},
  \bibinfo{journal}{Physical Review} \textbf{\bibinfo{volume}{40}},
  \bibinfo{pages}{749} (\bibinfo{year}{1932}).

\bibitem[{\citenamefont{Ville}(1948)}]{Ville:1948}
\bibinfo{author}{\bibfnamefont{J.}~\bibnamefont{Ville}},
  \bibinfo{journal}{C{\^a}bles et Transmission} \textbf{\bibinfo{volume}{2A}},
  \bibinfo{pages}{61} (\bibinfo{year}{1948}).

\bibitem[{\citenamefont{Martin and Flandrin}(1985)}]{MartinFlandrin:1985}
\bibinfo{author}{\bibfnamefont{W.}~\bibnamefont{Martin}} \bibnamefont{and}
  \bibinfo{author}{\bibfnamefont{P.}~\bibnamefont{Flandrin}},
  \bibinfo{journal}{IEEE Transactions on Acoustics, Speech, and Signal
  Processing} \textbf{\bibinfo{volume}{33}}, \bibinfo{pages}{1461}
  (\bibinfo{year}{1985}).

\bibitem[{\citenamefont{Lu and
  Vaswani}(2009)}]{lu2009wienerkhinchintheoremnonwidesense}
\bibinfo{author}{\bibfnamefont{W.}~\bibnamefont{Lu}} \bibnamefont{and}
  \bibinfo{author}{\bibfnamefont{N.}~\bibnamefont{Vaswani}},
  \emph{\bibinfo{title}{The wiener-khinchin theorem for non-wide sense
  stationary random processes}} (\bibinfo{year}{2009}), \eprint{0904.0602},
  \urlprefix\url{https://arxiv.org/abs/0904.0602}.

\bibitem[{\citenamefont{Dechant and Lutz}(2015)}]{Dechant_2015}
\bibinfo{author}{\bibfnamefont{A.}~\bibnamefont{Dechant}} \bibnamefont{and}
  \bibinfo{author}{\bibfnamefont{E.}~\bibnamefont{Lutz}},
  \bibinfo{journal}{Physical Review Letters} \textbf{\bibinfo{volume}{115}}
  (\bibinfo{year}{2015}), ISSN \bibinfo{issn}{1079-7114},
  \urlprefix\url{http://dx.doi.org/10.1103/PhysRevLett.115.080603}.

\bibitem[{\citenamefont{Daubechies et~al.}(1991)\citenamefont{Daubechies,
  Jaffard, and Journe}}]{Daubechies:1991wv}
\bibinfo{author}{\bibfnamefont{I.}~\bibnamefont{Daubechies}},
  \bibinfo{author}{\bibfnamefont{S.}~\bibnamefont{Jaffard}}, \bibnamefont{and}
  \bibinfo{author}{\bibfnamefont{J.~L.} \bibnamefont{Journe}},
  \bibinfo{journal}{SIAM J. Math. Anal.} \textbf{\bibinfo{volume}{22}},
  \bibinfo{pages}{554} (\bibinfo{year}{1991}).

\bibitem[{\citenamefont{Necula et~al.}(2012)\citenamefont{Necula, Klimenko, and
  Mitselmakher}}]{Necula_2012}
\bibinfo{author}{\bibfnamefont{V.}~\bibnamefont{Necula}},
  \bibinfo{author}{\bibfnamefont{S.}~\bibnamefont{Klimenko}}, \bibnamefont{and}
  \bibinfo{author}{\bibfnamefont{G.}~\bibnamefont{Mitselmakher}},
  \bibinfo{journal}{Journal of Physics: Conference Series}
  \textbf{\bibinfo{volume}{363}}, \bibinfo{pages}{012032}
  (\bibinfo{year}{2012}),
  \urlprefix\url{https://doi.org/10.1088%2F1742-6596%2F363%2F1%2F012032}.

\bibitem[{\citenamefont{Cornish}(2020)}]{Cornish:2020odn}
\bibinfo{author}{\bibfnamefont{N.~J.} \bibnamefont{Cornish}}
  (\bibinfo{year}{2020}), \eprint{2009.00043}.

\bibitem[{\citenamefont{Johnson et~al.}(2026)\citenamefont{Johnson,
  Chatziioannou, and Summers}}]{Johnson:2026rrn}
\bibinfo{author}{\bibfnamefont{A.}~\bibnamefont{Johnson}},
  \bibinfo{author}{\bibfnamefont{K.}~\bibnamefont{Chatziioannou}},
  \bibnamefont{and} \bibinfo{author}{\bibfnamefont{J.}~\bibnamefont{Summers}}
  (\bibinfo{year}{2026}), \eprint{2606.21473}.

\bibitem[{\citenamefont{Vajpeyi et~al.}(2026)\citenamefont{Vajpeyi, Mentasti,
  Baghi, Burke, and Speri}}]{Vajpeyi:2026msr}
\bibinfo{author}{\bibfnamefont{A.}~\bibnamefont{Vajpeyi}},
  \bibinfo{author}{\bibfnamefont{G.}~\bibnamefont{Mentasti}},
  \bibinfo{author}{\bibfnamefont{Q.}~\bibnamefont{Baghi}},
  \bibinfo{author}{\bibfnamefont{O.}~\bibnamefont{Burke}}, \bibnamefont{and}
  \bibinfo{author}{\bibfnamefont{L.}~\bibnamefont{Speri}}
  (\bibinfo{year}{2026}), \eprint{2606.20269}.

\bibitem[{\citenamefont{Talbot et~al.}(2021)\citenamefont{Talbot, Thrane,
  Biscoveanu, and Smith}}]{Talbot:2021igi}
\bibinfo{author}{\bibfnamefont{C.}~\bibnamefont{Talbot}},
  \bibinfo{author}{\bibfnamefont{E.}~\bibnamefont{Thrane}},
  \bibinfo{author}{\bibfnamefont{S.}~\bibnamefont{Biscoveanu}},
  \bibnamefont{and} \bibinfo{author}{\bibfnamefont{R.}~\bibnamefont{Smith}},
  \bibinfo{journal}{Phys. Rev. Res.} \textbf{\bibinfo{volume}{3}},
  \bibinfo{pages}{043049} (\bibinfo{year}{2021}), \eprint{2106.13785}.

\bibitem[{\citenamefont{Talbot et~al.}(2025)}]{Talbot:2025vth}
\bibinfo{author}{\bibfnamefont{C.}~\bibnamefont{Talbot}} \bibnamefont{et~al.},
  \bibinfo{journal}{Class. Quant. Grav.} \textbf{\bibinfo{volume}{42}},
  \bibinfo{pages}{235023} (\bibinfo{year}{2025}), \eprint{2508.11091}.

\bibitem[{\citenamefont{Schmidt}(1907)}]{Schmidt1907}
\bibinfo{author}{\bibfnamefont{E.}~\bibnamefont{Schmidt}},
  \bibinfo{journal}{Mathematische Annalen} \textbf{\bibinfo{volume}{63}},
  \bibinfo{pages}{433} (\bibinfo{year}{1907}),
  \urlprefix\url{http://eudml.org/doc/158296}.

\bibitem[{\citenamefont{Cornish}(2025)}]{Cornish:2025awt}
\bibinfo{author}{\bibfnamefont{N.~J.} \bibnamefont{Cornish}}
  (\bibinfo{year}{2025}), \eprint{2511.10632}.

\bibitem[{\citenamefont{Pearson and Cornish}(2026)}]{Pearson:2025wfd}
\bibinfo{author}{\bibfnamefont{N.}~\bibnamefont{Pearson}} \bibnamefont{and}
  \bibinfo{author}{\bibfnamefont{N.~J.} \bibnamefont{Cornish}},
  \bibinfo{journal}{Phys. Rev. D} \textbf{\bibinfo{volume}{113}},
  \bibinfo{pages}{064033} (\bibinfo{year}{2026}), \eprint{2509.05479}.

\end{thebibliography}

\end{document}